\documentstyle[11pt,titlepage,twoside]{article}

\newcommand{\sect}[1]{\setcounter{equation}{0}\section{#1}}
\newcommand{\subsect}[1]{\subsection{#1}}

\newtheorem{theo}{{\bf Theorem}}[section]
\newtheorem{prop}{{\bf Proposition}}[section]
\newtheorem{exam}{{\bf Example}}[section]

\newtheorem{defi}{{\bf Definition}}[section]

\newcommand{\beq}{\begin{equation}}
\newcommand{\eeq}{\end{equation}}
\newcommand{\benum}{\begin{enumerate}}
\newcommand{\eenum}{\end{enumerate}}
\newcommand{\beqn}{\begin{eqnarray}}
\newcommand{\eeqn}{\end{eqnarray}}
\newcommand{\beqnn}{\begin{eqnarray*}}
\newcommand{\eeqnn}{\end{eqnarray*}}

\newcommand{\down}{\nabla}
\newcommand{\pa}{\partial}

\newcommand{\calA}{{\mathcal{A}}}
\newcommand{\calF}{{\mathcal{F}}}
\newcommand{\calG}{{\mathcal{G}}}
\newcommand{\calH}{{\mathcal{H}}}
\newcommand{\calI}{{\mathcal{I}}}
\newcommand{\calP}{{\mathcal{P}}}
\newcommand{\calS}{{\mathcal{S}}}
\newcommand{\calV}{{\mathcal{V}}}

\newcommand{\bfB}{{\bf B}}
\newcommand{\bfF}{{\bf F}}
\newcommand{\bfJ}{{\bf J}}
\newcommand{\bfR}{{\bf R}}
\newcommand{\bfS}{{\bf S}}

\newcommand{\bfa}{{\bf a}}
\newcommand{\bfb}{{\bf b}}
\newcommand{\bff}{{\bf f}}
\newcommand{\bfg}{{\bf g}}
\newcommand{\bfn}{{\bf n}}
\newcommand{\bfp}{{\bf p}}
\newcommand{\bfu}{{\bf u}}

\newcommand{\bfx}{{\bf x}}
\newcommand{\bfy}{{\bf y}}
\newcommand{\bfz}{{\bf z}}

\begin{document}

\begin{titlepage}
\begin{center}

{\huge {\bf On the Dirac Approach to Constrained Dissipative
Dynamics}}
\vspace*{2.5cm} \\
{\large Sonnet Q. H. Nguyen} $^\flat$ $^\star$,
{\large {\L}ukasz A. Turski} $^\flat$ $^\sharp$\\
$^\flat$ Center for Theoretical Physics, Polish Academy of
Sciences,
Al. Lotnik{\' o}w 32/46, 02-668 Warsaw, Poland.\\
$^\star$ Institute of Mathematics, Polish Academy of Sciences,
{\' S}niadeckich 8, P.O. box 137, 00-950 Warsaw, Poland.\\
$^\sharp$ College of Sciences,
Al. Lotnik{\' o}w 32/46, 02-668 Warsaw, Poland.\\
\vspace{0.2cm} E-mails: ~ {\it sonnet@cft.edu.pl} ~ and ~ {\it
laturski@cft.edu.pl}
\end{center}
\begin{abstract}

In this paper we propose a novel algebraic and geometric
description for the dissipative dynamics. Our formulation bears
some similarity with the Poisson structure for non-dissipative
systems. We develop a canonical description for constrained
dissipative systems through an extension of the Dirac brackets
concept, and we present a new formula for calculating Dirac
brackets. This formula is particularly useful in description of
dynamical systems with many second-class constraints. After
presenting necessary formal background we illustrate our method
on several examples taken from particle dynamics, continuum
media physics and wave mechanics.

\end{abstract}
\end{titlepage}


\sect{Introduction}

The first systematic attempt to provide mathematically
consistent quantization procedure for constrained systems was
given by P.M. Dirac \cite{Dirac1}, who derived a formal
``replacement'' for the canonical Poisson brackets which plays
today a fundamental role in the canonical formalism for
constrained Hamiltonian systems on both classical and quantum
levels.  In spite of the considerable attention paid to that
formula in the mathematical literature
\cite{Marsden,Bhaskara,Sudarshan}, and several attempts to use
the Dirac brackets in the quantization of gauge invariant
systems, eg. \cite{Deriglazov,Ferrari} etc., until recently
there were few attempts to actually use this formalizm in more
conventional applications. We have recently provided few
examples of these applications in  classical and continuum
mechanics \cite{SN-LAT2}.

The canonical formalizm applies to the conservative systems.
These systems form a relatively small sub-class of interesting
physical systems since most of the others describing phenomena
at some effective rather than fundamental level are dissipative.
In the past several attempts were done to  describe dissipative
classical mechanics in a fashion similar to its canonical
description. One of these attempts, so called metriplectic
approach \cite{LAT,Morrison} was advocated as a natural
extension of the \textit{mixed canonical-dissipative dynamic}
proposed by  Enz \cite{Enz0}. In this paper we extend further
the concept of metriplectic dynamics to what we shall call {\it
semimetric-Poissonian dynamics}, which is nothing but a natural
combination of semimetric dynamics (a dissipative part) and
Poissonian dynamics (a conservative part). We propose a
canonical description for constrained dissipative systems
through an extension of the concept of Dirac brackets
\cite{Dirac1} developed originally for conservative constrained
Hamiltonian dynamics, to the non-Hamiltonian, namely {\it
metric} and mixed {\it metriplectic}, constrained dynamics. It
turns out that this generalized unified formula for the Dirac
brackets is very useful in the description and analysis of a
wider class of dynamical systems.   To proceed with our approach
we develop a new formula for calculating Dirac brackets which is
particularly effective in finding equations of motion and
constants of motions for systems with many constraints.

In order to make this paper self-contained we include in
Section~II a short ``primer~'' to the Poisson geometry, dynamics
and the Dirac brackets.

The rest of the paper is organized into four sections. In
Section~III we discuss semimetric dynamics and some elementary
features of the related mathematical structures like {\it
semimetric algebras} and {\it SJ-identity} the late should be
regarded as dynamical symmetric version of the Jacobi identity.
We develop symmetric concepts in the analogy with those in the
Poisson category in Appendix.

In Section~IV we discuss semimetric dynamics subject to some
constraints. We derive a symmetric analogue of the Dirac brackets
and provide its geometric interpretation  as induced metric on
some submanifold of a Riemannian manifold.   We also present a
new effective algorithm for calculation of the Dirac brackets in
both symmetric and antisymmetric cases.   Few examples, for
finite and infinite dimensional cases, are also discussed in
some detail.

In the section V we discuss semimetric-Poissonian dynamics which
is a combination of semimetric dynamics, discussed in the
section III, and Poissonian dynamics mentioned in the section
II.   This section also includes the extended Dirac approach to
constrained semimetric-Poissonian which is a combination of the
Dirac approach to constrained semimetric dynamics discussed in
the section IV and the usual to constrained Poissonian dynamics.

In  Section~VI we discuss interesting physical examples:  the
dissipative formulation of the Schr{\"o}dinger equation due to
Gisin \cite{Gisin}, the Landau-Lifshitz-Gilbert equations for
damped spins \cite{LAT,LAT-Holyst}, dynamics of damped rigid body
-- including among other the Morrison equation of damped rigid
body \cite{Morrison}. We also discuss a novel description of
the incompressible viscous fluid.

Some important but not crucial mathematical aspects of the
extended canonical formalism for constrained metric and mixed
semimetric-Poissonian systems will be given in the forthcoming
publication \cite{SN4}. Computational aspects of the Dirac
brackets -- symmetric, antisymmetric or mixed -- will be publish
shortly \cite{SN5}.

\sect{Poisson geometry, dynamics and Dirac brackets}

From the algebraic point of view the {\it Poisson algebra} is a linear
space $\cal F$ equipped with two structures:
\begin{description}
\item{{\bf i})}~ the commutative algebra structure with the
                 (associate) multiplication
                 $\calF \times \calF \to \calF$,
                 the product of two elements $f,g$ is
                 denoted simply by $fg$,
\item{{\bf ii})}~ the Lie algebra structure with the Lie bracket
                 $\{\cdot,\cdot\}: \calF \times \calF \to \calF$,
                 the product of two elements $f,g$ is
                 denoted by $\{f,g\}$,
\end{description}
related to each other by the Leibniz rule:
\begin{equation}
  \{fg,h\} = f\{g,h\} + \{f,h\}g \,.
\end{equation}
The Lie bracket of a Poisson algebra is called the {\it Poisson
bracket}. \\

The {\it Poisson manifold} is a smooth manifold $M$ for which
the commutative algebra of smooth functions on $M$,
$C^\infty(M)$, is equipped  with the Poisson bracket.  The
Poisson bracket $\{\cdot,\cdot\}$ acts on each function as a
derivation, thus there exists a contravariant $(2,0)$-tensor
$\Pi$ such that $\{f,g\}=\Pi(df,dg)$ for every functions $f,g$.
In the local coordinates $(z^k)$
\begin{equation} \label{Poisson.tensor}
  \{f,g\}(z) = \sum_{i,j}^m
  \Pi^{ij}(z) ~ \pa_i f ~ \pa_j g ~,~
  \pa_k \equiv \frac{\partial}{\partial z^k}\;.
\end{equation}
The tensor $\Pi$ which defines a Poisson bracket is called a
{\it Poisson tensor}. The antisymmetry of Poisson bracket
implies that tensor $\Pi$ must be antisymmetric, so $\Pi^{ij} =
- \Pi^{ji}$. The Jacobi identity requires that
\begin{equation} \label{jacobi-equi}
   \sum_{l=1}^{N}
   \Pi^{li} ~\partial_l \Pi^{jk} +
   \Pi^{lj} ~\partial_l \Pi^{ki} +
   \Pi^{lk} ~\partial_l \Pi^{ij} = 0 \,.
\end{equation}
A {\it Hamiltonian vector field} generated by a function $h$ is
a vector field defined by $X_h(f) = \{f,h\}$ for $\forall f$.
All flows generated by Hamiltonian vector fields-- {\it
Hamiltonian flows}, preserve the Poisson structure.

{\it Poisson dynamics} or {\it generalized Hamiltonian dynamics}
is a dynamics generated by some Hamiltonian vector field, for
which the Hamiltonian function plays physically specific role.

Let $TM$ and $T^*M$ denotes the tangent, and cotangent bundle of
the manifold $M$.  The Poisson tensor $\Pi$ induces a bundle map
$\Pi^\sharp: T^*M \to TM$ which is defined by $\Pi^\sharp(df) :=
X_f$ for all functions $f$. The {\it rank} of a Poisson
structure at point $z$ is defined to be the rank of
$\Pi^\sharp_z: T^*_zM \to T_zM$, which is equal to the rank of
matrix $\Pi^{ij}(z)$ in the local coordinates $(z^k)$. A Poisson
structure with constant rank equal to the dimension of the
manifold $M$ is called {\it symplectic} or {\it nondegenerate}.
In this case the ``inverse map'' of $\Pi$, denoted by $\omega$,
is a symplectic $2$-form and $\omega(X_f,X_g) = \{f,g\} =
\Pi(df,dg)$. The Darboux theorem states that  for every
symplectic structure there exists,  locally, a canonical
coordinates system $(x_1,\ldots,x_k,p_1,\ldots,p_k)$ such that
$\Pi = \frac{\pa}{\pa \bfx} \wedge \frac{\pa}{\pa \bfp}$,
($\omega=d\bfx \wedge d\bfp$), or equivalently,
$\{x_i,x_j\}=\{p_i,p_j\}=0$, $\{x_i,p_j\}=\delta_{ij}$. \\
The invariance of the Poisson structure under Hamiltonian flows
implies the constancy the tensor $\Pi$ rank along the orbits of
such flows. The orbit of each point of $M$ under the action of
all Hamiltonian flows forms a symplectic manifold  called a {\it
symplectic leave}. Since $M$ is a union of such orbits every
Poisson manifold is a smooth union of disjoint connected
symplectic manifold (symplectic leaves) of various ranks. \\
The splitting theorem \cite{Weinstein} for Poisson manifolds,
states that locally every Poisson manifold is the product of a
symplectic manifold and a Poisson manifold with zero rank.  In
the other words, locally in the neighborhood of the point
$\bfz_0$ there always exist canonical coordinates system:\\
$(x_1,\ldots,x_k,p_1,\ldots,p_k, z_{2k+1}, \ldots, z_n)$ such
that $\{x_i,x_j\}=\{p_i,p_j\}=0, ~ \{x_i,p_j\} = \delta_{ij}, ~
\{x_i,z_l\}=\{p_j,z_l\}=0, ~ \{z_r,z_s\}
= A_{rs}$ and $A_{rs}(\bfz_0)=0$. \\
The map $\varphi: M_1 \to M_2$ between two Poisson manifolds is called
{\it Poisson mapping} iff  $\{f \circ \varphi, g \circ \varphi \}_1 =
\{f,g\}_2 \circ \varphi$. The  {\it Poisson mapping} is a natural
generalization of the well-known  from classical mechanics notion of
the {\it canonical transformation}. \\
In the usual formulation of the classical mechanics the
constrained dynamics can be visualized geometrically as the
dynamics on some submanifold of the system phase space.
Similarly, the constrained Poisson dynamics can be represented
as such on some submanifold of the Poisson manifold.  However,
it is not always  possible to define induced Poisson structure
on a submanifold and therefore we have no obvious way how to
generalize constrained Poisson dynamics.  If the Poisson
structure is non-degenerate (symplectic case) then on each
submanifold there exists an induced two form which becomes
symplectic if it is non-degenerate.  Further for an arbitrary
submanifold $N$ of a symplectic manifold $(M,\omega)$ there
always exists a maximal submanifold $N' \subset N$ such
$\omega_{|N'}$ is non-degenerate,
then $N'$ has an induced symplectic hence Poisson structure. \\
Dirac \cite{Dirac1} has proposed an algebraic procedure to deal
with constrained dynamics.  First consider any finite set of
(linearly independent) constraints 
$\calA = \{a_1,\ldots,a_k\} \subset \calF \equiv
C^\infty(M)$ and define weakly-vanishing (or weakly-zero)
elements as linear combinations of constraints with arbitrary
functions coefficients, i.e. $\calI = \{ \sum_i f_i a_i, ~
\mbox{ where } ~f_i \in \calF,a_i \in \calA \}$. \\
Element $f \in \calF$ is called {\it first-class} with respect
to the set of constraints $\calA$ iff it has weekly-zero bracket
with all constraints, i.e. $\forall a\in \calA : ~ \{f, a\} \in
\calI$. Otherwise, it is called {\it second-class}. The set of
all first-class elements denoted by $\calF_1(\calA)$, forms a
linear subspace of $\calF$ and the set of all second-class
elements, denoted by $\calF_2(\calA)$.  \\
This classification divides the set of constraints $\calA$ into two
subsets: first-class constraints $\calA_1=\calA \cap
\calF_1(\calA)$ and second-class constraints $\calA_2=\calA \cap
\calF_2(\calA)$.  The number of second-class constraints must be even 
${\calA}_2=\{\Theta_1,\ldots,\Theta_{2s} \}$.  Dirac has
proven the Gramm matrix of second-class constraints,
$[\{\Theta_i,\Theta_j\}] = W$, is weakly non-degenerate. This
allowed him to define new antisymmetric Leibniz bracket, known
as the {\it Dirac bracket}:
\begin{eqnarray}\label{DirMain}
    \{f,g\}_D &=& \{f,g\} - \sum_{i,j=1}^{2s} \{f,\Theta_i\} C_{ij}
                           \{\Theta_j,g\}
\end{eqnarray}
where $C=[C_{ij}]=W^{-1}$ is an inverse matrix of ~$W$. Using
(\ref{DirMain}) one can  check that the Dirac brackets posses
all the required properties of the Poisson brackets. The
(algebraic) proof of the Jacobi identity is difficult.  One can
easily check that all the second-class constraints are Casimirs
with respect to the Dirac bracket, i.e. $\{\Theta_k,f\}_D=0$ for
all $f$.


\sect{Semimetric manifolds and semimetric dynamical systems}

In this section we introduce a concept of {\it semimetric
algebras} and {\it semimetric manifolds} which play a similar
role in the description of dissipative systems as the Poisson
algebras and the Poisson manifolds in the description of
conservative dynamics.  In the last two sections of this work we
will show that the {\it semimetric structure} together with the
{\it Poisson structure} are sufficient for  ``canonical''
description of a wide class of dissipative dynamical
systems. \\

Let $X$ be non-empty set.  A {\it semimetric bracket} on the
linear space of real functions defined on $X$, namely
$\calF=Fun(X)$, is a bilinear operation $\prec\cdot,\cdot\succ:
\calF \times \calF \to \calF$ which satisfies the following
requirements:
\begin{description}
\item{{\bf i)}}~
      It is symmetric:  $\forall f,g \in \calF ~:~ \prec f,g \succ
                   =\prec g,f \succ$.
\item{{\bf ii)}}~ It satisfies the Leibniz rule: $\forall f,g,h \in \calF ~:~
           \prec f g,h \succ = \prec f,h \succ g +
              f \prec g,h \succ$.
\item{{\bf iii)}}~ It is non-negative definite:
           $\forall f \in \calF ~:~ \prec f,f\succ \geq 0$, i.e. the
           function $\prec f,f\succ$ is non-negative definite function,
           $\forall x \in X: \prec f, f \succ(x) \geq 0$.
\end{description}
Note that if a  bilinear operation on $\calF$ satisfies
conditions {\bf i)}, {\bf ii)} it is called {\it pseudo-metric
bracket} or {\it symmetric Leibniz bracket}.   A symmetric
Leibniz bracket which satisfies the condition
\begin{description}
\item{{\bf iii*)}} Positive definite:
      $\forall f \in \calF ~:~ \prec f,f \succ \geq 0$ and
      $\prec f,f \succ = 0$ iff $f=\mathrm{Const}$ (at least locally),
\end{description}
is called {\it metric bracket}.

{\it Symmetric Leibniz algebra} is a linear space $\calF$
equipped with two structures: commutative algebra structure with
the (associate, commutative) multiplication and symmetric
(non-associate, non-commutative) structure and these two
structures are related by the Leibniz rule. {\it Semimetric
algebra} is a symmetric Leibniz algebra whose bracket
is semimetric. \\
As we shall see the semimetric algebra can be used to describe
the (wide class) of the dissipative  classical systems akin to
the description of the non-dissipative dynamics by means of the
Poisson
algebra. \\
From now we assume that $X$ is a smooth finite dimensional manifold,
namely {\it Phase space}, and $\calF=C^\infty(X)$ is a space of
all smooth functions on $X$. \\
There is a correspondence\footnote{Note that it is true for
smooth functions $C^\infty(X)$, but false for $C^k(X)$.}  one to
one between the symmetric Leibniz brackets on the space of
functions and the symmetric tensors on the manifold $X$: $\prec
f,g \succ = G(df,dg)$ where $G$ is a contravariant tensor field
of the type $(2,0)$ on $X$.  In the local coordinates $(z^k)$,
each symmetric tensor is of the form $G(\bfz) = \sum_{i,j}
G^{ij}(\bfz) \frac{\pa}{\pa z_i} \otimes \frac{\pa}{\pa z_j}$,
where $G^{ij}=G^{ji}$, hence each symmetric Leibniz bracket
locally must be of the form:
\beqn \label{gen.metric}
  \prec f,g \succ(z) &=&
  \sum_{i,j=1}^N  ~ G^{ij}(z) ~ \frac{\pa f}{\pa z^i} ~
            \frac{\pa g}{\pa z^j} ~,~ f, g \in C^\infty(X) \;.
\eeqn
A symmetric bracket becomes semimetric bracket iff it is
non-negative,
i.e. the matrix $[G^{ij}]$ is non-negative definite. \\
A semimetric bracket is called a {\it metric bracket} if $G$ is
positive definite (non-negative and non-degenerate), i.e. with
constant maximal rank, $\mbox{rank} ~G = \mbox{dim} ~X$. In the
general the tensor $G$ may have a nonconstant rank which depends
on points.
\begin{defi}
{\em {\it Semimetric manifold} is a smooth manifold $M$ for
which the
     commutative algebra of smooth functions on $M$ is a semimetric
     algebra, i.e. it is equipped with a semimetric bracket.
     Geometrically, {\it semimetric manifold} can be viewed as a
     pair $(M,G)$ where $G$ is {\it semimetric tensor}
     (or {\it cometric tensor}), i.e. symmetric,
     non-negative definite: $G(df,df) \geq 0$ for every function $f$,
     contravariant $(2,0)$-tensor field.
}
\end{defi}
\begin{exam}
{\em  Tensor $G(\bfz) = \sum_{i=1}^k \frac{\pa}{\pa z_i} \otimes
                \frac{\pa}{\pa z_i} -
                \sum_{j=k+1}^n \frac{\pa}{\pa z_j} \otimes
                \frac{\pa}{\pa z_j}$ is symmetric and non-degenerate,
      but it is not non-negative definite.
      Tensor $G(\bfz) = \sum_{i=1}^k \frac{\pa}{\pa z_i} \otimes
                \frac{\pa}{\pa z_i}$ for $k<n$, and tensor
      $G(\bfz) = z_1^2 \frac{\pa}{\pa z_1} \otimes
                \frac{\pa}{\pa z_1} + (z_2^2 z_3^2)
                \frac{\pa}{\pa z_2} \otimes
                \frac{\pa}{\pa z_2} +
                \sum_{i=3}^n z_i^2  \frac{\pa}{\pa z_i} \otimes
                \frac{\pa}{\pa z_i}$
      are non-negative definite, but degenerate. ~ $\Box$
}
\end{exam}
A {\it dissipative vector field} generated by the function $h$
is a vector field defined by $X_h^D (f) = \prec f,h\succ$ for
all functions $f$. The flow generated by dissipative vector
field should be called a {\it dissipative flow}. Dissipative
flows essentially differ from the Hamiltonian counterpart: 
they do not preserve the symmetric structure.\\
If the $(2,0)$-tensor $G$ is positive-definite, there exists
symmetric $(0,2)$-tensor $\calG$, its inverse, such that
$G(df,dh)=\calG(X_f^D,X_h^D)$, which is exactly a Riemannian
metric tensor.  In the local coordinates $(z_k)$, if: $G(\bfz)=
\sum_{i,j} G^{ij}(\bfz) \frac{\pa}{\pa z_i} \otimes
\frac{\pa}{\pa z_j}$, the tensor $\calG(\bfz)= \sum_{i,j}
G_{ij}(\bfz) d z_i \otimes d z_j$
where $\sum_j G^{ij} G_{jk}= \delta^i_k$.\\
The concept of semimetric manifold then is a natural generalization 
concept of Riemann manifold.  It is analogous to a generalization
from symplectic manifold to the Poisson manifold.  \\
Similarly as in the Poissonian case, it is not always  possible
to define induced semimetric structure on a submanifold, hence
there is no obvious way to describe canonically constrained
semimetric dynamics.

The map $F: M_1 \to M_2$ between two semimetric manifolds is
called {\it semimetric mapping} iff it maps the  semimetric
structures, i.e. $\prec f \circ F, g \circ F \succ_1 = \prec
f,g\succ_2 \circ F$.
\begin{prop}  \label{lemma_gramma1}
{\em  Let $(\calF,\cdot,\prec \cdot,\cdot \succ)$ be semimetric
algebra.
\begin{description}
\item{{\bf a})}~
     The Schwartz inequality holds
     \beqn
           \forall f,g \in \calF : ~
         \prec f,f\succ\prec g,g\succ \geq \prec f,g\succ^2 \,.
     \eeqn
\item{{\bf b})}~
    Let $f_1, f_2, \ldots, f_n$ be arbitrary elements of $\calF$.
     Then the square matrix
     \beqn
        Gram(f_1,\ldots,f_n) &=&
        \left[
        \begin{array}{ccc}
            \prec f_1,f_1 \succ & \ldots & \prec f_1,f_{n} \succ  \\
            \prec f_2,f_1 \succ & \ldots & \prec f_2,f_{n} \succ  \\
            \ldots      &    \ldots    &    \ldots   \\
            \prec f_{n},f_1 \succ & \ldots & \prec f_{n},f_{n} \succ \\
        \end{array} \right]
     \eeqn
     is non-negative definite.   In particular,
     $\det Gram(f_1,\ldots,f_n) \geq 0$.
     Furthermore, if $\calF$ is a metric algebra, then
     $\det Gram(f_1,\ldots,f_n) = 0$ iff $\{f_i\}_{i=1}^n$ are
     affine linear dependent.
\end{description}
}
\end{prop}
{\bf Proof.} ~ {\bf a})~ Indeed, for each real number $\lambda
\in \bfR$ the expression $0 \leq \prec f-\lambda g, f- \lambda
g\succ = \lambda^2 \prec g,g\succ - 2 \lambda \prec  f,g\succ
+\prec f,f\succ$ is non-negative quadratic form in the real
number $\lambda$. Hence the discriminant $\triangle = 4\prec
f,g\succ^2 - 4 \prec f,f\succ \prec
g,g\succ\leq 0$ must be non-negative. \\
If $\calF$ is metric algebra, then $\prec f,g\succ^2 = \prec
f,f\succ\prec g,g\succ$ iff $f,g$ are affine linear dependent,
i.e. $f - \lambda g = \mathrm{Const}$. \\

{\bf b})~ For each vector $\bfa = (a_1,a_2,\ldots,a_n) \in
\bfR^n$, denote $\bfa \cdot \bff = \sum_i a_i f_i$, we have
$\bfa^T ~ Gram(f_1,\ldots,f_n) ~ \bfa = \prec \bfa \cdot \bff,
\bfa \cdot \bff \succ \geq 0$.

\begin{flushright}
{\bf Q.E.D.}
\end{flushright}
\begin{exam} \label{semi_Hilbert01} \

{\em
\benum
\item The natural Euclidean metric of $n$-dim Euclidean space $X=\bfR^n$
      induces a natural metric structure on $C^\infty(X)$
      \beqn\label{euclid_metric01}
           \prec f,g \succ (\bfx) = \sum_{i=1}^n \frac{\pa f}{\pa x_i}
                         \frac{\pa g}{\pa x_i},
           ~\mbox{ where } ~ f,g \in C^\infty(\bfR^n), ~ \bfx \in \bfR^n\,.
      \eeqn
      We shall call the metric bracket (\ref{euclid_metric01}) the
      Euclidean metric bracket.
\item Here is a simple but general construction of the semimetric
      structure in the space of smooth functionals over a Hilbert
      space.  Let $X$ be a Hilbert space with a scalar product
      $<\cdot \mid \cdot>$ and $A$ be a linear operator on $X$.
      One can define a semimetric structure on the space of all
      smooth functionals over $X$ as following
      \beqn \label{hilbert_metric_st01}
          \prec \Phi,\Psi \succ(\bff) &=&
          \left< A \frac{\delta \Phi}{\delta \bff}
           \mid A \frac{\delta \Psi}{\delta \bff} \right> \,.
      \eeqn
      For instance, let $X$ be a Hilbert space of functions
      $X = \{ \bff: \bfR^n \to \bfR^d\}$ with a scalar product
      \beqn
         (\bff|\bfg) = \sum_{i,j=1}^d \int d^n \bfx ~
          f_i (\bfx) G^{ij}(\bfx) g_j (\bfx) \,, ~ \mbox{ where }
          \nonumber\\
          \bff = (f_1,f_2,\ldots,f_d), ~ \bfg = (g_1,g_2,\ldots,g_d)
          \in X \,.
      \eeqn
      The above scalar product on $X$ defines a semimetric bracket on the
      space of smooth functionals over $X$ by
\begin{eqnarray}\label{functional_metric01}
  \prec\Phi,\Psi\succ(\bff) = \sum_{i,j=1}^d \int d^n \bfx ~
          \frac{\delta \Phi}{\delta f_i(\bfx)} G^{ij}(\bfx)
         \frac{\delta \Psi}{\delta f_j(\bfx)},
\end{eqnarray}
where $\Phi,\Psi \in C^\infty(X)$.\\

     To be specific let $d=n$ and $G$ be a differential operator of the 
     form $D^+ D$, for instance, let
     $G^{ij}(\bfx)= -[ a \frac{\pa^2}{\pa x_i \pa x_j} +
                    b \delta_{ij} \triangle ]$, where $a,b \geq 0$
     where $\triangle$ denotes the Laplace operator. The symmetric
     bracket (\ref{functional_metric01}) assume now the form
\begin{eqnarray}
   \prec\Phi,\Psi\succ(\bff) =  \int d^n \bfx  \left\{ a
        \left[\down\cdot \frac{\delta\Phi}{\delta\bff(\bfx)}\right]
        \left[\down\cdot \frac{\delta\Psi}{\delta\bff(\bfx)}\right]\,
        \right. \nonumber\\
        + \, b \sum_{i} \left.
          \left[ \down \frac{\delta\Phi}{\delta f_i(\bfx)} \right] \cdot
          \left[ \down \frac{\delta\Psi}{\delta f_i(\bfx)} \right]
        \right \}
\end{eqnarray}
As we shall see this is exactly a semimetric bracket need in
viscous fluid dynamics.  ~$\Box$
\eenum }
\end{exam}
We shall call {\it semimetric dynamics} a dynamics which is
governed by some semimetric (dissipative) vector field. In local
coordinate system $(z_k)$, we have the following system of
first-order differential equations:
\beqn
   \dot{z}^i = \prec z^i, \calS \succ = X_{\calS}^D(z^i) = \sum_{j=1}^N
    G^{ij}(z) \frac{\pa \calS}{\pa z^j} \,, ~ i, j= 1,2,\ldots, N,
\eeqn
where $\prec \cdot,\cdot \succ$ is semimetric bracket,
$X_{\calS}^D$ is a dissipative vector field generated by
function $\calS$ defined on the phase space, and $G$ is a
semimetric tensor. In some very special cases the function
$\calS$ has a physical interpretation as entropy. $\calS$ is
always non-decreasing, since
\beqn
    \dot{\calS} = \prec \calS, \calS \succ \geq 0 \,.
\eeqn
If $f,g$ are Constants for semimetric dynamics, then the Leibniz
rule ensures that $fg$ is  also  a constant, but $\prec f,g
\succ$ is usually not a constant due to the lack of the Jacobi
identity. We shall now introduce a new concept a symmetric
analogy of the Jacobi Identity which we call  the {\it
SJ-identity}.
\begin{description}
\item{{\bf iv})} ~ {\it  SJ-identity}:
\beqn  \label{symmetric-Jacobi1}
   \forall ~ f,g,h \in \calF: ~
   2 \prec \prec f,g \succ, h \succ &=&
       \left[ \prec \prec f, h \succ, g \succ +
              \prec \prec g, h \succ, f \succ  \right] \nonumber\\
    ~ \Longleftrightarrow
   \forall~ f,h \in \calF: ~ \prec \prec f,f \succ, h\succ
                         &=& \prec \prec f,h \succ, f \succ \,.
\eeqn
\end{description}
A symmetric dynamical system is called a {\it SP-dynamics}
(``symmetric Poisson dynamics'') iff the symmetric Jacobi
identity (\ref{symmetric-Jacobi1}) holds.
\begin{prop}
{\em If $f,g$ are Constants of the SP-dynamics, then  $\prec f,g
\succ$ also is a Constant of motion.   In particular, if $f$ is
a constant, then $\prec f,f \succ$ also. }
\end{prop}
{\bf Proof.}~  Indeed, since $\prec f, \calS \succ= \prec g,
\calS \succ=0$ we have \beqn
  \frac{d}{dt} \prec f,g \succ =
  \prec \prec f,g \succ, \calS \succ = \frac{1}{2}
  \left[  \prec \prec f, \calS \succ, g \succ
        + \prec \prec g, \calS \succ, f \succ
  \right] =0 \,.
\eeqn
\begin{flushright}
{\bf Q.E.D.}
\end{flushright}
Therefore constants of SP-dynamics form a subalgebra. This property
is quite useful in finding constants of motion for SP-dynamics.\\
Note that each autonomous dynamical system described by the
system of the first order differential equations
\beqn\label{autonomic_pseudo01}
         \dot{x}_k = F_k(x_1,\ldots,x_n) \,, ~k=1,\ldots,n,
\eeqn
is a pseudo-metric system. Indeed, one can choose as the
function $\calS$, $\calS=\sum_{k=1}^n x_k$, and as a diagonal
pseudo-metric
$G^{ij}(x) = \delta^{ij} F_i(x)$. \\
Furthermore, locally and almost everywhere each dynamical system
(\ref{autonomic_pseudo01}) is metric.   Indeed, for each point
from the set $\{ \bfx ~: ~\forall k=1,\ldots,n: F_k(\bfx) \neq 0
\}$ there exists a neighborhood $U$ such that functions $F_k$ do
not change their sign inside $U$.  Let denote $s_k = \mbox{sign}
F_k = \pm 1$ in $U$. In the neighborhood $U$, the system should
be regarded as metric system with, for instance, $G^{ij}(x) =
s_i \delta^{ij} F_i(x) \geq 0$ and $\calS = \sum_k s_k x_k$.  In
particular, locally and almost everywhere, Poisson dynamics also
admit a metric description.
Each dynamical system of the type $\dot{x}_k =F_k(\bfx)$ should
be regarded as Poissonian system after doubling the number of
variables.  Indeed, consider a canonical Poisson structure:
$\{x_i,x_j\}=0=\{p_i,p_j\}, ~ \{x_i,p_j\}=\delta_{ij}$ and let
$\calH(\bfx,\bfp) = \sum_k p_k F_k(\bfx)$, the canonical
equations follow
\beqn
    \dot{x}_k = \{x_k, \calH \} = F_k(\bfx), ~
    \dot{p}_k = \{p_k, \calH \} =
        - \sum_j p_j \frac{\pa F_j(\bfx)}{\pa x_k} \,.
\eeqn
A system is {\it non-Poissonian} (or {\it non-metric}) if it can
not be written in the Poisson (resp. metric) form without
changing the number of variables. Each dynamical system of the
type $\dot{x}_k =F_k(\bfx)$ where functions $F_k(\bfx)$ are
positive-definite, is a metric system but it is (in the general)
non-Poissonian.  The answer to the question which Poissonian
system admits a global metric description,
remains unknown. \\
Morrison \cite{Morrison} has pointed out that metric dynamical
systems admit an asymptotic stability at isolated maxima of
function $\calS$. To show that let $\bfx$ be an isolated maximum
of $\calS$, then certainly $\pa_i \calS =0$ at $\bfx$, hence
$\bfx$ is an equilibrium point of the semimetric dynamical
system: $\dot{z}^i = \prec z^i, \calS \succ = \sum_j
G^{ij}(\bfz) \pa_j \calS$.  Define the function $L(\bfz)=
\calS(\bfz)- \calS(\bfx)$, then obviously $L(\bfx)=0$ and
$L(\bfz) < 0$ in some neighborhood of $\bfx$, since $\bfx$ is
the isolated maximum.   Furthermore, $\dot{L}(\bfz)= \prec L,
\calS \succ(\bfz)= \prec \calS, \calS \succ(\bfz) \geq 0$ for
$\bfz \neq \bfx$ and $\dot{L}(\bfx)=0$, therefore $L$ is Liapunov
function for the system and $\bfx$ is its stable equilibrium
point.   Now if the system is metric, then $\dot{L}(\bfz)= \prec
\calS, \calS \succ(\bfz) > 0$ for $\bfz \neq \bfx$ since
function $\calS$ is not locally constant.   Hence $\bfx$ is
asymptotically stable point. Generalization of the above
construction to the infinite dimensional case is not known.


\sect{Constrained metric dynamics}

In the frame work of symplectic geometry, constrained
Hamiltonian dynamics can be represented by a triplet
$(M,N,\omega)$ where $(M,\omega)$ is a symplectic manifold,
namely Phase space, and $N$ is a constraint submanifold of $M$.
Antisymmetric Dirac bracket for second-class constraints
\cite{Dirac1,Sniatycki} is nothing but the Poisson bracket on
some symplectic manifold $N' \subset N$, called  the
second-class
constraint manifold \cite{Sniatycki} et. al. (also in \cite{Marsden}). \\
Similarly, constrained metric dynamics should be represented by
a triplet $(M,N,g)$ where $g$ is metric tensor which is
responsible for a dissipation and $N$ is a constraint
submanifold of $M$.  We show that symmetric Dirac bracket for a
triple $(M,N,g)$ is nothing but the semimetric bracket on the
submanifold $N$.  It is worthy to note that any submanifold of a
Riemannian manifold is second-class with respect
to the metric bracket defined by the metric tensor. \\
Suppose that we have a pair $(M,\xi)$ where $M$ is a smooth
manifold and $\xi$ is non-degenerate symmetric $(0,2)$-tensor on
$M$.  Then at each point $x \in M$ the map $\xi_x: T_x M \times
T_x M \to R$ is bilinear, symmetric and non-degenerate and  it
induces a linear bijection $\xi_x^\sharp: T_x^*M \to T_xM$.
Further, if $\xi$ is Riemann metric tensor, then $\xi_x$ is a
scalar product on $T_xM$. The non-degeneracy of the tensor $\xi$
guarantees an existence of a
$(2,0)$-tensor field $\Lambda: T^*M \times T^*M \to R$.\\
Let $N$ be a submanifold of $M$ and $\xi_{|N}$ is supposed to be
non-degenerate.  Then at each point $x$ of the submanifold $N$,
the linear space $T_xM$ decomposes into direct sum of a tangent
space to $N$, $T_x N$, and its orthogonal with respect to
bilinear symmetric functional $\xi_x$, i.e. $T_xM=T_xN \oplus
(T_xN)^\perp$. The symmetric Dirac bracket with respect to a
triple $(M,N,\xi)$ is defined by
\beqn
    \Lambda_D(\alpha, \beta) &=&
    \xi( P \xi^\sharp(\alpha), P \xi^\sharp(\beta)) \,, ~
    \mbox{ where $P$ is a projection} \nonumber\\
    & & \mbox{ onto $T_xN$ along $(T_xN)^\perp$, $\alpha, \beta$ are
    $1$-forms on $M$ } .
\eeqn
The symmetric Dirac bracket in the space of functions is then
\beqn
    \prec f, g \succ_D &=& \Lambda_D (df,dg) \,, ~
    \forall f,g \in C^\infty(M).
\eeqn
If $N$ is the second-class submanifold
\beqn
   N &=& \{ x \in M : ~ \Theta_i (x) = 0 \} \,,
\eeqn
then we derive explicite formula for symmetric Dirac bracket
with respect to the triple $(M,N,\xi)$.   Let  us denote $X_g$ a
vector field generated by the function $g$, i.e. $X_g(f) =
\Lambda(df,dg) = \prec f,g \succ = \xi(X_f,X_g)$.   Let
$W=[W_{ij}] = [ \prec \Theta_i, \Theta_j \succ ] = [  \xi(
X_{\Theta_i}, X_{\Theta_j} ) ]$ and $C=[C_{ij}]=W^{-1}$. It is
easy to see that the vector fields $X_{\Theta_i}$ span
$TN^\perp$, then orthogonal projection $Q$ onto $TN^\perp$ along
$TN$ has a form:
\beqn
    Q X = \sum_{i,j} \xi(X, X_{\Theta_i}) C_{ij} X_{\Theta_j} \,.
\eeqn
Then the orthogonal projection $P$ onto $TN$ along $TN^\perp$ is
of the form $PX = X -QX$,  hence we have $P X_f = X_f -
\sum_{i,j} \xi(X, X_{\Theta_i}) C_{ij} X_{\Theta_j}$. Then the
symmetric Dirac formula is of the form
\beqn \label{dirac_sym_br001}
   \prec f,g \succ_D &=& \Lambda_D(df,dg) =
                     \xi(P X_f, P X_g) = \xi(X_f, P X_g)
    \nonumber\\
    &=& \xi(X_f,X_g) - \xi(X_f,Q X_g)  \nonumber\\
    &=& \prec f,g \succ - \sum_{i,j} \prec f,\Theta_i \succ C_{ij}
                                     \prec \Theta_j,g \succ \,,
\eeqn
which coincides  with the antisymmetric Dirac bracket formula
(\ref{DirMain}) for Poisson bracket with the antisymmetric
brackets $\{\cdot,\cdot\}$ replaced by $\prec \cdot, \cdot\succ$. 
Our procedure shown above applies  to both cases, for symmetric or 
antisymmetric $(0,2)$-tensors.\\
The Dirac formula (\ref{dirac_sym_br001}) plays a key role in
the practical use of the Dirac brackets.    Algebraically, one
may use it as a definition of the Dirac bracket for an arbitrary
symmetric or antisymmetric algebra.  The disadvantage of the
algebraic approach is based on the fact that it is very
difficult to understand why the Jacobi identity for the new
Dirac bracket holds when the above procedure is applied to a
Poisson bracket.  In the metric context, if the algebraic
formula (\ref{dirac_sym_br001}) is regarded as a definition of
the Dirac bracket, then it is easy to check that new algebraic
Dirac bracket is symmetric Leibniz bracket (i.e. the algebraic
properties {\bf i)}, {\bf ii)} hold), but the crucial
non-negativity property {\bf iii)} is not easy to verify.   A
simple proof will be given latter after the Theorem
{\bf\ref{theorem_new_formu01}}.  ~ It is easy to see that
\beqn
   \prec \Theta_a, f \succ_D = 0 \,,
   \mbox { for arbitrary function } f(z) \,,
\eeqn
therefore all constraints $\Theta_a$ are Casimirs, i.e. belong
to the Centrum of the semimetric algebra
$(C^\infty(X),\prec\cdot,\cdot\succ_D)$. \\
Now we would like to present the new formula for calculating
symmetric or antisymmetric Dirac bracket.
\begin{theo} \cite{SN5} \label{theorem_new_formu01}\

{\em The following identity holds:
\beqn \label{twodet}
   \prec f,g \succ_D &=& \frac{\det W_{f,g} }{\det W} \,, ~ \forall f,g
   \in \calF \,,
\eeqn 
where 
\beqn\label{twodetW}
    W = \left[
        \begin{array}{ccc}
            \prec\Theta_1,\Theta_1\succ & \ldots &
            \prec\Theta_1,\Theta_{N}\succ \\
            \prec\Theta_2,\Theta_1\succ & \ldots &
            \prec\Theta_2,\Theta_{N}\succ \\
            \ldots      &    \ldots    &    \ldots \\
            \prec\Theta_{N},\Theta_1\succ & \ldots &
            \prec\Theta_{N},\Theta_{N}\succ
        \end{array} \right] , ~~\nonumber\\
    W_{f,g} = \left[
        \begin{array}{cccc}
            \prec\Theta_1,\Theta_1\succ & \ldots &
            \prec\Theta_1,\Theta_{N}\succ & \prec\Theta_1,g\succ \\
            \prec\Theta_2,\Theta_1\succ & \ldots &
            \prec\Theta_2,\Theta_{N}\succ & \prec\Theta_2,g\succ\\
            \ldots      &    \ldots    &    \ldots   &  \ldots \\
            \prec\Theta_{N},\Theta_1\succ & \ldots &
            \prec\Theta_{N},\Theta_{N}\succ & \prec\Theta_N,g\succ \\
            \prec f,\Theta_1\succ & \ldots &
            \prec f,\Theta_{N}\succ & \prec f,g\succ
        \end{array} \right] .
\eeqn
The same formula holds for antisymmetric Dirac brackets. }
\end{theo}
The proof is straightforward; one  applies twice the Laplace
recursive formula for the determinant expansion to the last
column and row of the matrix $W_{f,g}$.  ~$\Box$ \\

One consequence of (\ref{twodet}) and (\ref{twodetW}) for semimetric 
bracket is that $\forall f$ we have $\prec f,f \succ_D = \frac{\det
W_{f,f}}{\det W}$, and therefore the inequality $\prec f,f \succ_D 
\geq 0$ holds, according to the Proposition {\bf \ref{lemma_gramma1}}.
In one constraint case, this is equivalent to the Schwartz inequality. \\
The theorem {\bf \ref{theorem_new_formu01}} provides a new
effective formula for calculating Dirac brackets for both
symmetric and antisymmetric case. Usually, the direct attempt to
use the formula (\ref{dirac_sym_br001}) is impractical for a
system with rather big number of  constraints of second type.
This is  because it requires quite complicated evaluation of the
elements of the inverse matrix $C_{ij}$. Note, however, that
when we are not interested in the Dirac brackets but only in
resulting equations for constrained dynamics then the evaluation
of $C=W^{-1}$ is unnecessary. Applying theorem {\bf
\ref{theorem_new_formu01}} we find immediately equation of the
quantity $f$ in constrained symmetric/antisymmetric dynamics:
\beqn
    \dot{f} = \prec f,\calH \succ_D = \frac{\det W_{f,\calH} }{\det W} \,,~
    \mbox{ or } ~
    \dot{f} = \{f,\calH\}_D = \frac{\det W_{f,\calH} }{\det W} \,.
\eeqn
This formula is particularly convenient for finding constants of
motion  for constrained -- symmetric and antisymmetric --
dynamics. Indeed,  $f$ is a constant of motion  for constrained
dynamics iff ~ $\det W_{f,\calH} =0$. \\
Below we show few examples of the constrained symmetric Dirac
brackets which are applicable in differential geometry and
physics.
\begin{exam} \label{sfere01}
{\em  Consider the standard Euclidean metrics
      $\prec z^i,z^j \succ = \delta^{ij} = G^{ij}$  and the fixed surface
      $f(\bfz)=0$ in $\bfR^n$.
      Here the function $f$ is assumed to be smooth and with zero as it
      regular value, that is $f^{-1}(0)$
      is a close regular $n-1$ dimensional differential
      submanifold in $\bfR^n$. Using $f(\bfz)=0$ as a constraint we
      find the Dirac semimetric brackets
      \beqn
        \prec z^i, z^j \succ_D = \delta^{ij} - n^i n^j  \,,
      \eeqn
      where ${\bf n}(\bfz) = \frac{\down f}{||\down f||}$ is an unit
      normal vector to the surface at $\bfz$.  The metric tensor
      $G_D^{ij}(\bfz) = \delta^{ij} - n^i n^j$ is
      nothing but the induced metric tensor on this surface. ~ $\Box$
}
\end{exam}
The following example illustrates how to derive the metric
structure for lattice-spins \cite{LAT}, that is a set of
classical spins $\vec S_a$ where $a$ labels the lattice sites
\begin{exam} \label{spin-ex1}
{\em  We introduce the usual  lattice-spin metric brackets as
      \beqn    \label{metric-spin01}
         \prec S^i_a,S^j_b \succ &=& \delta_{ab}
                         \delta^{ij} |S_a| = G^{ij}_{ab}, ~
                         i,j=1,2,3, ~ a,b = 1,2,\ldots,N\,,
      \eeqn
      and we define $2N$-dim surface $\calP$ by the following system
      of $N$ constraints
      \beqn
         \Theta_a(\bfS) = |S_a|^2 - r_a^2 = \sum_{i=1}^3 (S_a^i)^2 - r_a^2
         =0 \,, ~ a=1,2,\ldots,N.
      \eeqn
      The Dirac metric brackets for the surface $\calP$ are
      \beqn\label{metric-spin02}
         \prec S^i_a, S^j_b \succ_D =\delta^{ij}  |S_a|
          \left[\delta_{ab} - \frac{S^i_a S^j_b}{S_a^2} \right] \,.
      \eeqn
}
\begin{flushright}
$\Box$
\end{flushright}
\end{exam}
The next example illustrates the metric structure for energy-conserving 
rigid body.
\begin{exam} \label{rigid-ex1} \

{\em One may postulate the metric brackets for a rigid body
as \beqn \label{metric-rigid000}
   \prec \omega_i,\omega_j \succ &=& \delta_{ij} K(\omega)\,,
         \mbox{ where $K$ is some function of the rigid body} \nonumber\\
     & & \mbox{ angular frequency $\omega$} \,.
\eeqn
\begin{description}
\item{{\bfa})}~ Consider energy as constrained surface
    \beqn
         \Theta(\omega) = \sum_{k=1}^3 I_k \omega_k^2 - E \,.
    \eeqn
    Calculating the Dirac metric brackets one gets
    \beqn \label{metric-rigid001}
        \prec \omega_i,\omega_j \succ_{\{\Theta\}} = K(\omega)
        \left[ \delta_{ij}  - \frac{I_i I_j
              \omega_i \omega_j}{\sum_k I_k^2 \omega_k^2} \right]\,.
    \eeqn
    We can consider some particular cases:
    \begin{description}
    \item{{\bfa1})}~ One may choose $K=\sum_{k=1}^3 I_k^2 \omega_k^2$,
          where $I_k$ are moments of inertia with respect to main axes
          of the rigid body.  Then the Dirac metric brackets are
          \beqn \label{metric-rigid02}
              \prec \omega_i,\omega_j \succ_{\{\Theta\}} = \delta_{ij}
              \left[ \sum_{k=1}^3 I_k^2 \omega_k^2 \right]
              - I_i I_j \omega_i \omega_j \,.
          \eeqn
          This metric structure coincides with the metric structure
          postulated by Morrison \cite{Morrison}.
    \item{{\bfa2})}~ Since the Poisson structure of rigid body is
          the same as for classical spins, we may postulate $K=|\omega|$,
          where $|\omega|=\sqrt{\sum_{k=1}^3 \omega_k^2}$.
          The Dirac metric brackets follow
          \beqn \label{metric-rigid06}
               \prec \omega_i,\omega_j \succ_{\{\Theta\}}= |\omega|
               \left[ \delta_{ij} - \frac{ I_i I_j \omega_i \omega_j }
                             { \sum_{k=1}^3 I_k^2 \omega_k^2  }
               \right] \,.
          \eeqn
    \end{description}
\item{{\bfb})}~ Consider Poissonian Casimir as constrained surface
      \beqn
          \Theta(\omega) = \sum_{k=1}^3 \omega_k^2 - |\omega_0|^2 \,.
      \eeqn
      The Dirac metric brackets follow
      \beqn
         \prec \omega_i,\omega_j \succ_{\{\Theta\}}= K(\omega)
          \left[ \delta_{ij} - \frac{ \omega_i \omega_j }
                             { |\omega|^2  }
          \right] \,.
      \eeqn
      For instance, $K(\omega)=|\omega|$, we get
      \beqn \label{metric-rigid07}
          \prec \omega_i,\omega_j \succ_{\{\Theta\}}= |\omega|
              \left[ \delta_{ij} - \frac{ \omega_i \omega_j }
                             { |\omega|^2  }
              \right] \,.
      \eeqn
\end{description}
}
\begin{flushright}
$\Box$
\end{flushright}

\end{exam}

Next examples show how our formalizm works in the  Hilbert
spaces.
\begin{exam}
{\em  Let $Ph=L^2(\bfR^n; \bfR^d)$ be a Hilbert space of real,
vector valued
      square integrable functions  with standard scalar product
      $<\cdot \mid \cdot>$, i.e.
      $<\bff \mid \bfg> = \int d^n \bfx ~ \bff(\bfx) \cdot \bfg(\bfx)$.
      where $\bff = (f_1,\ldots,f_d), ~ \bfg = (g_1,\ldots,g_d).$
      Let $||\bff||^2 = <\bff \mid \bff>$.
      In the space of all smooth functionals over $Ph$, according to
      the construction given in the Example {\bf \ref{semi_Hilbert01}},
      the semimetric structure may be defined by
      \beqn\label{L2_hilbert}
         \prec \phi_1,\phi_2 \succ(\bff) = \sum_{i=1}^d
          \int d^n \bfx
            \left[ \frac{\delta \phi_1}{\delta f_i(\bfx)}
                   \frac{\delta \phi_2}{\delta f_i(\bfx)}
            \right] \,,~ \phi_1, \phi_2 : Ph \to R \,,
      \eeqn
      where $\frac{\delta \phi_i}{\delta f_k}$ denotes a Gateaux functional
      derivative.  \\
      Consider a surface of infinite dimensional sphere $S^{\infty}$
      with radius $r$ as a subspace with one constraint
      \beqn
         S^\infty = \{  \bff \in Ph ~:~  ||\bff||^2 = r^2 \} \,.
      \eeqn
      Now we calculate the Dirac metric structure for the sphere
      $S^{\infty}$.
      The metric bracket (\ref{L2_hilbert}) can be rewritten,
      introduce the {\it canonical metric tensor} $G$
      \beqn
         G^{ij} (\bfx,\bfy) = \prec f_i(\bfx),f_j(\bfy) \succ =
         \delta_{ij} \delta(\bfx-\bfy) \,,
      \eeqn
      as
      \beqn  \label{funct_form1}
         \prec \phi_1, \phi_2 \succ(\bff) &=&
         \int d^n \bfx d^n \bfy \frac{1}{2}
         \left[
              \frac{\delta \phi_1}{\delta f_i(\bfx)}
              \frac{\delta \phi_2}{\delta f_j(\bfy)} +
              \frac{\delta \phi_1}{\delta f_j(\bfy)}
              \frac{\delta \phi_2}{\delta f_i(\bfx)} \right]
              \prec f_i(\bfx),f_j(\bfy) \succ \,. \nonumber\\
      \eeqn
      The Dirac semimetric structure on $S^\infty$ follows
      \beqn  \label{sfere001}
        \prec f_i(\bfx), f_j(\bfy) \succ_D &=& \delta_{ij}
         \delta(\bfx-\bfy) - \frac{f_i(\bfx) f_j(\bfy)}{||\bff||^2}  \,.
      \eeqn
}
\begin{flushright}
$\Box$
\end{flushright}
\end{exam}
\begin{exam} \label{metric-QM}
{\em  Let $\Psi = \Psi_1 + i \Psi_2$ and its complex conjugate
      $\Psi^* = \Psi_1 - i \Psi_2$ where $\Psi_1, \Psi_2$ are
      real functions integrated by square, i.e. they belong to
      the Hilbert space $L^2$.   We define metric structure by
      \beqn
        \prec \Psi_k(\bfx),\Psi_l(\bfy) \succ = \frac{1}{2}
        \delta_{kl} \delta(\bfx-\bfy),~ \mbox{ where } ~ k,l=1,2.
      \eeqn
      One can rewrite it to the form
      \beqn\label{metric-QM01}
         \prec \Psi(\bfx),\Psi(\bfy) \succ =
         \prec \Psi^*(\bfx),\Psi^*(\bfy) \succ = 0, ~
         \prec \Psi(\bfx),\Psi^*(\bfy) \succ = \delta(\bfx-\bfy)\,,
      \eeqn
      which we call by {\it canonical metric bracket for Quantum
      Mechanics}.
      The Dirac structure on the sphere $||\Psi||=Const$ in the Hilbert
      space  follows
      \beqn  \label{metric-QM02}
         \prec \Psi(\bfx),\Psi(\bfy) \succ_D =
           - \frac{\Psi(\bfx) \Psi(\bfy)}{2 || \Psi ||^2 }, ~
         \prec \Psi^*(\bfx),\Psi^*(\bfy) \succ_D =
           - \frac{\Psi^*(\bfx) \Psi^*(\bfy)}{2 || \Psi ||^2 }  \nonumber\\
         \prec \Psi(\bfx),\Psi^*(\bfy) \succ_D = \delta(\bfx-\bfy)
           - \frac{\Psi(\bfx) \Psi^*(\bfy)}{2 || \Psi ||^2 }\,.
      \eeqn
      The Dirac brackets of the components $\Psi_k$ one get in the
      accordance with (\ref{sfere001})
      \beqn
        \prec \Psi_k(\bfx), \Psi_l(\bfy) \succ_D = \frac{1}{2}
        \left[ \delta_{kl}  \delta(\bfx-\bfy) -
        \frac{\Psi_k(\bfx) \Psi_l(\bfy)}{ ||\Psi||^2} \right ]  \,.
      \eeqn
}
\begin{flushright}
$\Box$
\end{flushright}
\end{exam}
\begin{exam} \label{metric_viscous_fl}
{\em  (The canonical description for incompressible, viscous fluid
       dynamics is based on this example)\\
      Let $Ph=W^{(1,2)}(\bfR^n; \bfR^n)$ be a Sobolev space of real
      functions.
      In the space of all smooth functionals over $Ph$ we introduce
      a semimetric structure
      \beqn \label{L2_sobolev}
          \prec \phi_1,\phi_2 \succ(\bfJ) = \int d^n \bfx  \left\{ a
        \left[\down\cdot \frac{\delta \phi_1}{\delta\bfJ(\bfx)}\right]
        \left[\down\cdot \frac{\delta \phi_2}{\delta\bfJ(\bfx)}\right]
        + b \sum_{i}
        \down \left[\frac{\delta \phi_1}{\delta J_i(\bfx)} \right] \cdot
        \down \left[\frac{\delta \phi_2}{\delta J_i(\bfx)}\right]
        \right\}  \,, \nonumber\\
      \eeqn
      where $\mathbf{J}$ denotes  the real vector in the n-dimensional
      Euclidean space, and $a,b$ are real non-negative coefficients.
      We can rewrite this semimetric structure in the form
      \beqn
           \prec J_i(\bfx),J_j(\bfy) \succ &=&
           - \left[ a \frac{\pa}{\pa x_i}
           \frac{\pa}{\pa x_j}  +  b \delta^{ij} \triangle \right]
           \delta(\bfx-\bfy) \,.
      \eeqn
      Consider infinite-dimensional subspace $\calV$ of
      divergence-free functions (incompressibility condition) as a
      system with infinite number of constraints:
      \beqn
         \calV = \{ \bfJ \in Ph ~:~ \forall \bfx ~~~
         \Theta_{\bfx}(\bfJ) = \down_{\bfx} \cdot \bfJ(\bfx) = 0 \} \,.
      \eeqn
      Then the Dirac semimetric structure for the subspace
      $\calV$ follows:
      \beqn  \label{J.incom01}
          \prec J_i(\bfx),J_j(\bfy) \succ_D &=& -
          \left[ a \frac{\pa}{\pa x_i} \frac{\pa}{\pa x_j}
                 +  b \delta^{ij} \triangle \right]
           \delta(\bfx-\bfy) - \nonumber\\
          & & \int d\bfz d\bfz'
           \prec J_i(\bfx), \Theta (\bfz) \succ  C(z,z')
           \prec \Theta(\bfz'), J_j(\bfy) \succ \,, \nonumber\\
      \eeqn
      where $C$ is an inverse symmetric operator of the constraint matrix
      \beqn  \label{J.incom02}
        C(\bfx,\bfy) &=& \frac{1}{a+b} \int d\bfz  G(|\bfx-\bfz|)
                           G(|\bfz-\bfy|) \,,
      \eeqn
      here $G(|\bfx-\bfy|)$ denotes the standard Green function
      (fundamental distribution) of the Laplace equation, i.e.
      $\triangle_{\bfx} G(|\bfx-\bfy|) = \delta(\bfx-\bfy)$.
      Putting eq. (\ref{J.incom02}) back to eq. (\ref{J.incom01})
      and after some simple calculations finally we obtain
      \beqn \label{J.incom03}
          \prec J_i(\bfx),J_j(\bfy) \succ_D  &=&
          -b \left[ \delta_{ij} \triangle_{\bfx} -
                   \frac{\pa^2}{\pa x_i \pa x_j} \right]
          \delta(\bfx-\bfy) \,.
      \eeqn
      Physically, the last eq. (\ref{J.incom03}) fully describes
      {\it dissipative structure for incompressible viscous
      fluid}. We shall see that more clearly in
      subsection \ref{subsect6.5}.
}
\begin{flushright}
$\Box$
\end{flushright}
\end{exam}


\sect{Semimetric-Poissonian systems}

The physical systems are usually dissipative.  It turns out that
both Poissonian and metric structures \textit{alone} are not
enough to describe dissipative systems.  However, a proper
combination of these two types of dynamics can, for many
interesting cases, provide a satisfactory and fully algebraic
description of dissipative dynamics.

A {\it semimetric-Poissonian bracket} on the space of functions
$\calF=C^\infty(X)$ is a bilinear operation
$\{\{\cdot,\cdot\}\}: \calF \times \calF \to \calF$ which is a
linear combination of a Poisson and a semimetric bracket
\beqn
  \forall f,g \in \calF ~:~ \{\{f,g\}\}=\{f,g\} - \prec f,g \succ \,,
\eeqn
where $\{\cdot,\cdot\}$ is the Poisson bracket and $\prec
\cdot,\cdot \succ$ is the semimetric bracket.
\begin{defi}
{\em {\it Semimetric-Poisson manifold} is a pair $(M,\Pi-G)$
     where $\Pi$ is a Poisson tensor, $G$ is a semimetric tensor.
}
\end{defi}
We shall call a {\it Semimetric-Poissonian dynamics} a dynamics
governed by the following system of equations:
\begin{eqnarray}\label{freeenergyW}
  \dot{z}^i = \{\{z^i, \Phi\}\} =
   \{z^i,\Phi\} - \prec z^i,\Phi \succ
      = X_\Phi(z^i) - X_\Phi^D(z^i) \,,
\end{eqnarray}
where $\Phi$ is some phase space function.   In real physical
applications it is often the case that  $\Phi$ has the interpretation
of the system free energy.  It is then a matter of convention to chose
minus sign in the eq.(\ref{freeenergyW}).  Indeed we have then
\begin{eqnarray}\label{DISSENER}
  \dot\Phi = \{\Phi,\Phi\}- \prec \Phi,\Phi \succ
  = - \prec \Phi,\Phi \succ \leq 0 \,.
\end{eqnarray}
what describes the dissipation of energy. \\
It is often even more convenient to  go a step further and
decompose function $\Phi$ into two parts: the internal
energy $\calH$ and the dissipation function $\calS$, so $\Phi=\calH-\calS$.
Hence, when $\calS$ is a Casimir of the Poissonian part,
$\{f,\calS\}=0$ for all $f$, the evolution of some ``observable'' $f$
in the semimetric-Poissonian dynamics follows
\beqn\label{SecondLaw}
   \dot{f} &=& \{\{ f, \Phi \}\} = \{f,\calH\} -
   \prec f, \calH - \calS \succ \,.
\eeqn
Equation (\ref{SecondLaw}) for $f=\calS$ give us the
semimetric-Poissonian formulation for the ``second law of
thermodynamics'', namely
\beqn\label{SecondLawII}
  \dot{\calS} + \prec \calS,\calH\succ &=&
  \prec \calS,\calS \succ \geq0\;,
\eeqn
where the expression on the l.h.s. of Eq.(\ref{SecondLawII}) is 
just the convective time derivative of $\calS$ along the time 
trajectory in the semimetric-Poissonian phase space.

Note that if $\bfx$ is an isolated minimum of $\Phi$, then the
function $L(\bfz) = \Phi(\bfz) - \Phi(\bfx)$ is a Liapunov
function for semimetric-Poisson system.  Hence, we obtain
\begin{prop}
{\em  If $\bfx$ is an isolated minimum of the free energy
function
      $\Phi$, then $\bfx$ is a stable equilibrium point for
      semimetric-Poissonian system
      $\dot{\bfz}=\{\bfz,\Phi\}- \prec \bfz,\Phi \succ$.
      Furthermore, if the system is metric-Poissonian, then
      $\bfx$ is asymptotically stable point.
}
\end{prop}
\begin{exam}
{\em Consider a modification of the harmonic oscillator
described by
     \beqn
         \dot{x}_1 =    x_2  - a x_1( x_1^2 + x_2^2 ), ~
         \dot{x}_2 = - x_1  - a x_2( x_1^2 + x_2^2 ) \,.
     \eeqn
     This system is semimetric-Poissonian with
     \beqn
         \{x_1,x_2\} = 1,  ~ ~
         \Phi = \frac{1}{2} \left[ x_1^2 + x_2^2 \right] \,,  \nonumber\\
         \prec x_1, x_1 \succ = a x_1^2, ~
         \prec x_2, x_2 \succ = a x_2^2, ~
         \prec x_1, x_2 \succ = a x_1 x_2 \,.
     \eeqn
     Point $(0,0)$ is an isolated minimum of $\Phi$, hence it is a
     stable equilibrium point for the system. ~~ $\Box$
}
\end{exam}
The proposed in preceding sections formal structure of
Poissonian and semimetric dynamics allows us to suggest the
following scheme for construction of  constrained dissipative
dynamics of real physical systems.
\begin{description}
\item{{\bf a})}~
     Consider a canonical Poisson structure and semimetric structure.
     The semimetric structure must be postulated according to our
     physical insight in the nature of relevant dissipative processes.
\item{{\bf b})}~
      Choose sets $A$, $B$ of second-class constraints for the Poissonian
      and semimetric part, respectively.   Note that any constraint is
      second-class for non-degenerate semimetric structure.
\item{{\bf c})}~
      Calculate antisymmetric and symmetric Dirac brackets with respect
      to the set of constraints $A$ and $B$ and later take a union of
      them.
\end{description}
Most interesting cases take place when the set of constraints
for the semimetric part is a subset of constants of motion for
the Poissonian part, i.e. system is dissipative however some of
constants of the Poisson dynamics remain constants for the
semimetric-Poissonian dynamics.  One may use this feature to
design many interesting dissipative systems.    As an example,
we illustrate how to design variety of damped rigid body
dynamics in the next section.


\sect{Applications: Physical Examples}


\subsect{Particle on the hypersurface with friction}
We shall begin our illustration of the Dirac bracket
applications by discussion of a simple example, namely the
classical particle moving with friction on a hypersurface
$\calS_{\mathrm{config}} =\{\bfx \in \bfR^n ~\mid ~f(\bfx)=0\}$.
Denote the particle position by $\bfx$ and its conjugate
momentum by $\bfp$ and let the friction force experienced by
that particle be proportional to the particle velocity. The
phase space  $\bfR^{2n}$ is now equipped with two structures:
canonical Poisson and semimetric structure.  Then the Poisson
structure follows  is described by 
\beqn
    \{x_i,x_j\}=0=\{p_i,p_j\}, ~ \{x_i,p_j\}=\delta_{ij} \,,
\eeqn 
and the semimetric structure is defined by
\beqn
    \prec x_i,x_j \succ=0=\prec x_i,p_j \succ, ~
    \prec p_i,p_j \succ= \delta_{ij} \lambda_i \,, ~
\eeqn
where $\lambda_i(\bfx) > 0$ is the directional and space
dependent damping coefficient. \\
The semimetric-Poisson structure is defined by
$\{\{\cdot,\cdot\}\}=\{\cdot,\cdot\}-\prec \cdot,\cdot \succ$. \\
Consider $\calH = \frac{\bfp^2}{2m} + V(\bfx)$, the dissipative
dynamics derived from the above structures follows
\beqn
     \dot{x}_i = \{\{x_i,\calH\}\} = \frac{p_i}{m}, ~
     \dot{p}_i = \{\{p_i,\calH\}\} = -\lambda_i(\bfx) \frac{p_i}{m} -
                 \frac{\pa V}{\pa x_i} \,,
\eeqn
which can be rewritten in the Newtonian form $m \ddot{x}_i +
\lambda_i(\bfx) \dot{x}_i - F_i(\bfx)=0$, where ${\bfF}(\bfx) =
-\frac{\pa V}{\pa \bfx}$ is a potential force. In the
particular, when $\lambda_i(\bfx)=\lambda(\bfx)$, we have $m
\ddot{\bfx}+\lambda(\bfx) \dot{\bfx} - \bfF(\bfx)=0$. \\
Next consider a fixed algebraic surface $f(\bfx)=0$ in $\bfR^n$.
We make assumptions that $f$ is smooth and zero is its regular
value.  The second assumption ensures that
$\calS_{\mbox{config}} = f^{-1}(0)$ is a close regular $n-1$
dimensional differential submanifold in $\bfR^n$.   Moreover
this assumption guarantees $\nabla f \neq 0$, so we can use the
gradient to define the normal vector on $\calS_{\mbox{config}}$.\\
The set of constraints consists now of two elements:
\beqn \label{type1}
  \Theta_1 & \equiv & f(\bfx) = 0\,, ~~
  \Theta_2 \equiv  \bfp \cdot \frac{\pa f}{\pa \bfx} = 0\,.
\eeqn
For the Poissonian dynamics both the above constraints
${\Theta_{i}}$ are
second-class in the Dirac classification. \\
Denoting the unit normal vector to the surface $f$ at the point
$x$ by $\bfn(x)$, $\bfn(\bfx)= \frac{1}{\left|\frac{\pa f}{\pa
\bfx } \right| } \frac{\pa f}{\pa \bfx} ~$, the antisymmetric
Dirac brackets for the Poissonian part of our construction are
\beqn
 \{ x_i,x_j \}_D &=& 0 \,, ~~
 \{ x_i,p_j \}_D  = \delta_{ij} -
    \frac{1}{\left|\frac{\pa f}{\pa \bfx }\right|^2}
    \frac{\pa f}{\pa x_i}\frac{\pa f}{\pa x_j}
     = \delta_{ij} - n_i (\bfx) n_j (\bfx) \,, \nonumber\\
 \{ p_i,p_j \}_D &=&
    \frac{1}{\left|\frac{\pa f}{\pa \bfx}\right|^2}
    \left\{\frac{\pa f}{\pa x_j}
           \left[\bfp \cdot\frac{\pa}{\pa\bfx}\right]
           \frac{\pa f}{\pa x_i} -
           \frac{\pa f}{\pa x_i}
           \left[\bfp \cdot\frac{\pa}{\pa\bfx}\right]
           \frac{\pa f}{\pa x_j}
    \right\} \nonumber \\
 &=& n_j(\bfx)
      \left[\bfp \cdot \frac{\pa}{\pa \bfx}\right]
     n_i(\bfx)
   - n_i(\bfx)
      \left[\bfp \cdot \frac{\pa}{\pa \bfx}\right]
     n_j(\bfx) \,.
\eeqn
For the semimetric dynamics only the second constraint
$\Theta_{2}$ is second-class.  The symmetric Dirac brackets for
the semimetric part
\beqn
  \prec x_i,x_j \succ_D = 0 = \prec x_i,p_j \succ_D \,, ~
  \prec p_i,p_j \succ_D = \delta_{ij} \lambda_{i} -
 \frac{ \lambda_i(\bfx) \lambda_j(\bfx)
       \frac{\pa f}{\pa x_i}\frac{\pa f}{\pa x_j} }
            {\sum_k \lambda_k(\bfx) \left|\frac{\pa f}{\pa x_k}\right|^2 }
   \,.
\eeqn
In particular, when $\lambda_i(\bfx)=\lambda(\bfx)$, the
symmetric Dirac brackets can be written in the form \beqn
  \prec x_i,x_j \succ_D = 0 = \prec x_i,p_j \succ_D \,, ~
  \prec p_i,p_j \succ_D = \lambda(\bfx)
       [\delta_{ij} - n_i(\bfx) n_j(\bfx)] \,.
\eeqn Finally take an union of these two Dirac structures
\beqn
 \{\{ x_i,x_j \}\}_D &=& 0 \,, ~~
 \{\{ x_i,p_j \}\}_D = \delta_{ij} - n_i (\bfx) n_j (\bfx) \,, \nonumber\\
 \{\{ p_i,p_j \}\}_D &=&
    \frac{1}{\left|\frac{\pa f}{\pa \bfx}\right|^2}
    \left\{\frac{\pa f}{\pa x_j}
           \left[\bfp \cdot\frac{\pa}{\pa\bfx}\right]
           \frac{\pa f}{\pa x_i} -
           \frac{\pa f}{\pa x_i}
           \left[\bfp \cdot\frac{\pa}{\pa\bfx}\right]
           \frac{\pa f}{\pa x_j}
    \right\}
    \nonumber \\
     & & - \lambda_{i}(\bfx) \left[ \delta_{ij}  -
   \frac{ \lambda_j(\bfx) \frac{\pa f}{\pa x_i}\frac{\pa f}{\pa x_j} }
        {\sum_k \lambda_k(\bfx) \left|\frac{\pa f}{\pa x_k}\right|^2 }
   \right]     \,.
\eeqn
When the Hamiltonian for that system has the form $\calH =
\frac{\bfp^2}{2m} + V(\bfx)$, the dissipative Hamilton-Dirac
equations of motion follow as
\beqn \label{mecha1}
  \dot{x}_i &=&  \{ x_i, \calH \} - \prec  x_i, \calH \succ
      = \frac{1}{m}\left[p_i - (\bfp \cdot \bfn) n_i \right]
      = \frac{p_i}{m} \,, \nonumber\\
 \dot{p}_i &=& \{ p_i, \calH \} - \prec  p_i, \calH \succ
    = F_i -\left[ \bfF \cdot \bfn + \frac{1}{m}\bfp \cdot
                       \left[\left(\bfp\cdot
                                   \frac{\pa}{\pa \bfx}
                             \right) \bfn \right]
           \right] n_i \, \nonumber\\
    & & - \frac{\lambda_i(\bfx)}{m}
    \left[ p_i - \frac{\pa f}{\pa x_i}
     \left(
        \frac{ \sum_{j} \lambda_j(\bfx)\frac{\pa f}{\pa x_j} p_j }
             {\sum_k \lambda_k(\bfx)
                     \left|\frac{\pa f}{\pa x_k}\right|^2 }
         \right)  \right] \,.
\eeqn
For isotropic damping, when $\lambda_i(\bfx)=\lambda(\bfx)$, we
can rewrite these equations of motion in the Newtonian form as
\beqn
     m ~ \ddot{\bfx} + \lambda(\bfx) \dot{\bfx} &=&
     \bfF(\bfx) - \left[ \bfF(\bfx) \cdot \bfn(\bfx) + m \dot{\bfx} \cdot
                           \frac{d}{dt}\bfn(\bfx) \right] \bfn(\bfx) \,.
\eeqn


\subsect{Variety dynamics of damped rigid body}

The usual Poisson brackets for a rigid body angular velocity
vector $\vec\omega$, after suitable rescaling, are
\beqn
    \{ \omega_i , \omega_j \} &=& 
    \sum_k \varepsilon_{ijk} \omega_k \,. 
\eeqn
Suppose that there is no second-class constraints for the Poisson
part of the rigid body dynamics and just one  such a constraint
for its symmetric part. Assume now that this constraint is such
that the system energy (or any Casimir function for the metric
part) is constant. The choice of that constraint determines
details of the semimetric-Poissonian structure. \\
Using the canonical metric brackets (\ref{metric-rigid001}) and
the constrained energy $\calH$ from example {\bf
\ref{rigid-ex1}} we obtain Dirac metric bracket for rigid
body (\ref{metric-rigid02}). Combining these two structures
one finds the metric-Poissonian brackets for rigid body \cite{Morrison}:
\begin{eqnarray}
   \{ \{ \omega_i , \omega_j \}\} &=& \{ \omega_i , \omega_j \} -
      \lambda \prec  \omega_i , \omega_j \succ_D  \nonumber \\
   &=& \varepsilon_{ijk} \omega_k  -  \lambda
   \left[ \delta_{ij} \left( \sum_{k=1}^3 I_k^2 \omega_k^2 \right)
        - I_i I_j \omega_i \omega_j \right] .
\end{eqnarray}
Consider the system ``free energy''
\beqn\label{energy_rigid01}
    \Phi= \calH - \calS =
    \frac{1}{2}\left[ \sum_{k=1}^3 I_k \omega_k^2 \right] -
    \calS(|\omega|^2)\,,
\eeqn
The equations of, energy conserving, motion for the damped
rigid body are
\beqn
    \dot{\omega}_1 &=& \{\{\omega_1, \Phi\}\} =
    \{ \omega_1,\calH \} + \lambda \prec \omega_1, \calS \succ_D \nonumber\\
    &=&  \omega_2 \omega_3 (I_2-I_3) + 2 \lambda \calS' \omega_1
    \left[ I_2(I_2-I_1) \omega_2^2 + I_3(I_3-I_1) \omega_3^2 \right] \,
    \nonumber\\
    & & \mbox{ and its cyclic permutation }\, .
\eeqn
The dissipation of the free energy follows
\beqn
   \dot{\Phi} &=& -\dot{\calS} = -\lambda \prec \calS, \calS \succ_D
   \nonumber\\
   &=& - \lambda
   \left[ \left|\frac{\pa \calS}{\pa \omega}\right|^2
          \left(\sum_{k=1}^3 I_k^2 \omega_k^2 \right) -
          \left( I_i \omega_i \frac{\pa \calS}{\pa \omega_i} \right)^2
   \right] \leq 0 \,,
\eeqn
and there is no dissipation of energy iff $I_1=I_2=I_3$. ~ $\Box$ \\
Alternatively, we can derive new semimetric-Poissonian bracket
by combining metric bracket (\ref{metric-rigid06}) with the
standard Poisson bracket for the rigid body
\beqn
   \{ \{ \omega_i , \omega_j \}\} &=& \{ \omega_i , \omega_j \} -
   \lambda \prec  \omega_i , \omega_j \succ_D \nonumber \\
   &=& \varepsilon_{ijk} \omega_k  -  \lambda |\omega|
   \left[ \delta_{ij} -
     \frac{ I_i I_j \omega_i \omega_j }{\sum_{k=1}^3 I_k^2 \omega_k^2}
   \right] . \nonumber\\
\eeqn
Equations of motion following from the free energy
(\ref{energy_rigid01}) are
\beqn
    \dot{\omega}_1 &=& \{\{\omega_1,\Phi\}\} =
    \{ \omega_1,\calH \} +\lambda \prec \omega_1, \calS \succ_D \nonumber\\
    &=&
    \omega_2\omega_3 (I_2-I_3) + 2 \lambda \calS' |\omega| \omega_1
    \frac{ I_2(I_2-I_1) \omega_2^2 + I_3(I_3-I_1) \omega_3^2 }
         { \sum_{k=1}^3 I_k^2 \omega_k^2 } \,
    \nonumber\\
    & & \mbox{ and its cyclic permutation }\,. ~ \Box
\eeqn
Finally we derive new semimetric-Poissonian bracket by combining
metric bracket (\ref{metric-rigid07}) with the standard Poisson
bracket for the rigid body
\beqn
   \{ \{ \omega_i , \omega_j \}\} &=& \{ \omega_i , \omega_j \} -
   \lambda \prec  \omega_i , \omega_j \succ_D \nonumber \\
   &=& \varepsilon_{ijk} \omega_k  -  \lambda |\omega|
   \left[ \delta_{ij} - \frac{ \omega_i \omega_j }{|\omega|^2}
   \right] .
\eeqn
Again using the free energy (\ref{energy_rigid01}) we find
\beqn
    \dot{\omega}_1 &=& \{\{\omega_1,\Phi\}\} = \{ \omega_1,\calH \}
       - \lambda \prec \omega_1, \calH \succ_D \nonumber\\
    &=&
    \omega_2\omega_3 (I_2-I_3) - \lambda |\omega| \omega_1
    \left[I_1-\frac{\sum_{k=1}^3 I_k \omega_k^2 }{ |\omega|^2}\right] \,
    \nonumber\\
    & & \mbox{ and its cyclic permutation }\, .
\eeqn
The above dynamics does not conserve the energy, but $|\omega|^2$ 
remains the system Casimir.  The energy dissipation is given by
\beqn
   \dot{\calH}= - \lambda \prec \calH,\calH \succ_D = - \lambda |\omega|
   \left[ \sum_k I_k^2 \omega_k^2 -
        \frac{ (\sum_k I_k \omega_k^2)^2}{|\omega|^2} \right] \leq 0 \,,
\eeqn
since $ \left(\sum_k I_k^2 \omega_k^2\right) |\omega|^2 \geq
\left( \sum_k I_k \omega_k^2 \right)^2 $. For $\lambda >0, ~
|\omega| > 0$, there is no dissipation of energy when
$I_1=I_2=I_3$.  ~$\Box$


\subsect{Classical damped spins}
The Poisson bracket for lattice-spins
\beqn
    \{ S_a^i , S_b^j \} &=& \delta_{ab} \varepsilon_{ijk} S_a^k \,,
\eeqn
where $a$ labels the spin location and $k=1,2,3$, guarantees
that length of each spin $|\bfS_a|$ is a Casimir. Our model is
based on the assumption that there are no second-class
constraints for the Poissonian part and only one constraint for
the metric part. \\
Starting from canonical metric bracket (\ref{metric-spin01}) and
using the length of the spins $|\bfS_a|$ constraint, as in the
example {\bf \ref{spin-ex1}}, we obtain the Dirac metric bracket
for lattice-spins (\ref{metric-spin02}). Combining these two
structures one finds the metric-Poissonian of lattice spins
\cite{LAT}:
\begin{eqnarray}\label{mp-spin01}
  \{\{ S_a^i , S_b^j \}\} &=& \{ S_a^i , S_b^j \} -
    \lambda \prec  S_a^i , S_b^j \succ_D\\  \nonumber
           &=& \delta_{ab} \varepsilon_{ijk} S_a^k - \lambda
    \delta^{ij}  |\bfS_a|
    \left[\delta_{ab} - \frac{S^i_a S^j_b}{S_a^2} \right] \,,
\end{eqnarray}
where $\lambda$ is the damping coefficient.  \\
Now we can easily derive the Landau-Lifshitz-Gilbert equation of
classical damped lattice-spins 
\beqn
   \dot{\bfS}_a &=&  \{\{ \bfS_a, \calH \}\} =
      \bfS_a \times \bfB_{ef,a} - \lambda
      \frac{\bfS_a \times ( \bfS_a \times \bfB_{ef,a} )}{|\bfS_a|} \,,
\eeqn 
where  $\bfB_{ef,a} = - \frac{\pa \calH}{\pa \bfS_a}$ is
the effective
magnetic field acting on the spin $\bfS_a$. \\
The dissipation of energy follows
\beqn
   \dot{\calH} = - \lambda \prec \calH,\calH\succ_D
   = - \lambda \sum_a |S_a| \left[ \bfB_{ef,a}^2 -
                     \frac{(\bfB_{ef,a}\cdot \bfS_a)^2 }{|\bfS_a|^2}
     \right] \leq 0 \,.
\eeqn


\subsect{Dissipative quantum mechanics}
Combining canonical Poisson brackets for quantum mechanics [QM]
\beqn
   \{ \Psi(\bfx), \Psi(\bfy) \} = \{ \Psi^*(\bfx), \Psi^*(\bfy) \}
   = 0, ~ \{ \Psi(\bfx), \Psi^*(\bfy) \} = \frac{1}{i\hbar}
   \delta(\bfx-\bfy) \,,
\eeqn
and constrained metric brackets for QM (\ref{metric-QM02}),
which were derived from the canonical metric bracket for QM
(\ref{metric-QM01}) for the physically important constraint,
namely the conservation of the wave function norm (probability
conservation) as in the example {\bf \ref{metric-QM}}, we find
{\it dissipative Metric-Poissonian structure for QM} :
\beqn
    \{\{ \Psi(\bfx), \Psi(\bfy) \}\} &=&
    \{ \Psi(\bfx), \Psi(\bfy) \} - \frac{\lambda}{\hbar}
    \prec \Psi(\bfx),\Psi(\bfy)\succ_D = \frac{\lambda}{\hbar}
    \frac{\Psi(\bfx) \Psi(\bfy)}{2 ||\Psi||^2}, \nonumber\\
    \{\{ \Psi^*(\bfx), \Psi^*(\bfy) \}\} &=&
    \{ \Psi^*(\bfx), \Psi^*(\bfy) \} - \frac{\lambda}{\hbar}
    \prec \Psi^*(\bfx),\Psi^*(\bfy)\succ_D = \frac{\lambda}{\hbar}
    \frac{\Psi^*(\bfx)\Psi^*(\bfy)}{2||\Psi||^2}, \nonumber\\
    \{\{ \Psi(\bfx), \Psi^*(\bfy) \}\} &=&
    \{ \Psi(\bfx), \Psi^*(\bfy) \} - \frac{\lambda}{\hbar}
    \prec \Psi(\bfx),\Psi^*(\bfy)\succ_D \nonumber\\
    &=& \frac{1}{i\hbar}\delta(\bfx-\bfy) - \frac{\lambda}{\hbar}
    \left[\delta(\bfx-\bfy)-
          \frac{\Psi(\bfx) \Psi^*(\bfy)}{2 || \Psi ||^2 } \right]\,,
\eeqn
where $\lambda$ is the (undefined) damping constant. \\
Using the conventional Hamiltonian for the Schr{\"o}dinger
equation
\beqn
   \calH(\Psi,\Psi^*) =
   <\Psi|H|\Psi> = \int d^n \bfx \Psi^*(\bfx) H \Psi(\bfx)\,,
\eeqn
where $H$ is the quantum mechanical (self-adjoined) operator we
obtain the  evolution of the wave function in the form:
\beqn \label{Gisin1}
    i \hbar \pa_t \Psi(\bfx) &=& i \hbar \{\{ \Psi(\bfx), \calH \}\} =
    H \Psi(\bfx) + i \lambda
    \left[ \frac{<\Psi|H|\Psi>}{||\Psi||^2} - H\right] \Psi(\bfx)\,.
\eeqn
This equation is known as the {\it Gisin dissipative wave
equation} \cite{Gisin}. Here the construction of the semimetric
bracket ensures that the norm of the state vector is reserved
(so the probability is conserved) since it is a second-class
constraint for the semimetric part. The dissipation of energy
follows
\beqn
    \dot{\calH} &=& -\prec \calH,\calH \succ = \frac{2\lambda}{\hbar}
         \left[ - ||H\Psi||^2 + \frac{<\Psi|H|\Psi>^2}{||\Psi||^2}\right]
    \leq 0 \,, \nonumber\\
     & & ~\mbox{(Schwartz inequality)} \,.
\eeqn
In the above equation the equality is achieved for  $\Psi$ which
are the eigenstates of the Hamiltonian. The damping in the Gisin
equation  refers, therefore, to the transition amplitudes only.
When the initial wave packet is constructed from eigenstates
corresponding to energies $E\geq E_0$ then the final state of
the evolution described by eq. (\ref{Gisin1}) is the eigenstate
with energy $E_0$. This property distinguishes  the Gisin
dissipative wave equation from other dissipative Schr{\"o}dinger
equations.



\subsect{Viscous fluid dynamics} \label{subsect6.5}

In the fluid mechanics the state of an isothermal fluid is
described by its  mass density and velocity fields
$(\varrho,\bfu)$ or by $(\varrho,\bfJ)$ where the  current ${\bf
J}$ field equals $\bfJ = \varrho \bfu$.  The semimetric
structure for fluid dynamics\cite{LAT} follows
\beqn \label{viscous_struc_diss01}
    \prec \varrho(\bfx),\varrho(\bfy) \succ = 0, ~
    \prec \varrho(\bfx),J_k(\bfy) \succ = 0, \nonumber\\
    \prec J_k(\bfx),J_l(\bfy) \succ = -
     \left[ a \frac{\pa^2}{\pa x_k \pa x_l} +
            \eta \delta_{kl} \triangle_x \right] \delta(\bfx-\bfy) \,,
\eeqn
where $a = \zeta + \frac{\eta}{3}$ and $\zeta,\eta$ are bulk and
shear viscosity respectively.  \\
The structure (\ref{viscous_struc_diss01}) is semimetric,
indeed $\forall \calG$ we have:
\beqn
   \prec \calG,\calG \succ(\varrho,\bfJ) &=& - \sum_{k,l} \int d\bfx d\bfy
    \frac{\delta \calG}{\delta J_k(\bfx)}
    \left\{ \left[ a \frac{\pa^2}{\pa x_k \pa x_l} +
                     \eta \delta_{kl} \triangle_x \right]
            \delta(\bfx-\bfy) \right\}
   \frac{\delta \calG}{\delta J_l(\bfy)}  \nonumber\\
   &=&  \int d\bfx
      \left\{  a \left[  \down \cdot
             \frac{\delta \calG}{\delta \bfJ(\bfx)} \right]^2 +
             \eta \sum_{k}
             \left| \down \left[\frac{\delta \calG}{\delta J_k(\bfx)}\right]
             \right|^2
      \right\}  \,.   \nonumber\\
\eeqn
Note, that this is not a metric bracket, since any functional
which depends only on $\varrho$, has zero semimetric bracket
with himself. Also note that the kinetic fluid energy $E_{kin} =
\int d\bfx ~ {\bfJ^2}/{2\varrho}$ dissipation follows directly
from (\ref{viscous_struc_diss01}):
\beqn \label{viscous_ene_diss01}
    \dot{E}_{kin} = - \prec E_{kin},E_{kin} \succ =
    - \int d\bfx
    \left\{  a \left[\down \cdot \frac{\bfJ}{\varrho(\bfx)} \right]^2 +
             \eta \sum_{k} \left| \down \frac{J_k}{\varrho(\bfx)}
                           \right|^2
    \right\}  \leq 0 \,. \nonumber\\
\eeqn
In classical hydrodynamics a particular role is played by the
incompressible fluid assumption, for incompressible viscous
fluid according to eq. (\ref{J.incom03}) we have
\beqn \label{incom-visc-dirac}
     \prec \varrho(\bfx),\varrho(\bfy) \succ_D = 0 \,, ~
     \prec \varrho(\bfx),J_i (\bfy) \succ_D = 0 \,, \nonumber\\
     \prec J_i(\bfx),J_j(\bfy) \succ_D  = -
          \eta \left[ \delta_{ij} \triangle_{\bfx} -
                   \frac{\pa^2}{\pa x_i \pa x_j} \right]
          \delta(\bfx-\bfy) \,.
\eeqn
The dissipative energy for viscous incompressible fluid should
be easily calculated
\beqn
   \dot{E}_{kin} &=& - \prec E_{kin},E_{kin} \succ_D =
   - \frac{\eta}{2} \int d\bfx
   \left\{  \sum_{k,l} \left( \frac{\pa u_k}{\pa x_l} +
                              \frac{\pa u_l}{\pa x_k} \right)^2
    \right\}  \leq 0 \,. \nonumber\\
\eeqn
Finally, put $\nu=\frac{\eta}{\varrho_0}$ the kinematic
viscosity, using (\ref{incom-visc-dirac}) and the Poisson part
calculated in \cite{SN-LAT2} we easily derive non-local
evolutional equations for incompressible viscous fluid
\beqn
    \frac{\pa \bfu}{\pa t} + (\bfu \cdot \down) \bfu - \nu
    \triangle \bfu &=& \down \int d\bfz ~ G(\bfx-\bfz) \down_{\bfz}
    \left[ (\bfu \cdot \down) \bfu - \nu
    \triangle_{\bfz} \bfu \right]  \,. \nonumber\\
\eeqn


\sect{Final comments}

The description of the dissipative systems dynamics is usually
undertaken within the framework of non-equilibrium statistical
mechanics. The ``simplified'' version of the full many-body
description often used in applications, for example in phase
transformations physics, statistical theory of turbulence,
granular media dynamics etc. is the kinetic equation  for the
time evolution of the ``relevant degrees of freedom''
distribution function. This equation is derived making strict
assumption  about the nature of  the fluctuations in the system
-- the underlying stochastic process performed by the relevant
system degrees of freedom in the full phase space of the
physical model. The equations of motion for the coarse grained
variables describing the meso- or macroscopic system properties
are obtained from that kinetic equation by one of the known
procedures, which once are well justified and understood, and
occasionally are just a heuristic chain of semi-mathematical
operations.

In this paper we have discussed a novel approach to description
of the dissipative systems dynamics, which is purely algebraic.
Instead of making series of assumptions on the level of
microscopic physics of the problem we assume existence of
certain algebraic structure, akin to that used in Hamiltonian
formulation of classical dynamics which permit us to derive
meso- or macroscopic dissipative equations directly from the
system free energy. The basic ingredient of that procedure, the
semimetric Leibniz bracket for dynamics variables over the whole
phase space, is postulated according to our knowledge about
dissipative processes.  With knowledge of the Poissonian
structure for non-dissipative part of the system dynamics, the
symmetries of the problem and using Dirac machinery
\cite{Dirac1} one can derive easily different Dirac structures,
which are otherwise hard  to postulated, describing dissipative
dynamics. One may use the algorithm proposed here to design
variety dissipative dynamical systems with required conservative
observables.

As shown in the paper this permits us to build up the
metriplectic dynamics scenario for several non-trivial systems:
classical particle physics, many spins dynamics, rigid body
dynamics, compressible and incompressible viscous fluid dynamics
and some quantum mechanical problems.  Several other
applications,  notably the relativistic charged particle systems
 which can also be formulated within the
metriplectic scenario have been discussed previously
\cite{IBB-LAT,LAT-ANK}. The quantum applications  are of
particular interest in view of some similarity between the
metriplectic approach and the Lindblads construction
\cite{Lindblad} -- the standard tool in dissipative quantum
system analysis. We expect to comment on the connection between
both these approaches in the forthcoming publication.

\begin{center}
{\LARGE Appendix}
\end{center}
A symmetric Leibniz bracket which satisfies the SJ-identity is
called the {\it SP-bracket}.   The tensor $G$ which generates
SJ-bracket is called {\it SP-tensor}.   In the local coordinates
\beqn
   \forall_{i,j = 1,2,\ldots,N}: ~  0 &=&
   \sum_{k=1}^N \left[ G^{kj} \frac{\pa G^{ii} }{\pa z^k} -
                       G^{ki} \frac{\pa G^{ij} }{\pa z^k}
                \right] \,.
\eeqn
Hence, SP-bracket and SP-tensor are symmetric analogy to Poisson
bracket and Poisson tensor, respectively.
In particular each symmetric tensor where $G_{ij}$ are constants
(i.e. do not depends on points), generates a SP-bracket.  For a
non-trivial example of SJ-tensor, see Example {\bf \ref{SP-tensor1}}.\\
{\it SP-manifold} is a pair $(M,G)$ where $G$ is a symmetric
     SP-tensor of the type $(2,0)$.  \\
\begin{exam} \label{SP-tensor1}
{\em It is easy to see that
     $G=\sum_{i,j=1}^N   z_i z_j  \frac{\pa}{\pa z_i} \otimes
            \frac{\pa}{\pa z_j}$ is SP-tensor.
     Hence, $(\bfR^N,G)$ is SP-manifold. ~ $\Box$
}
\end{exam}
Similarly, with the concept of SJ-identity one can define
SL-algebra, SP-algebra which are symmetric analogy of Lie
algebra and Poisson algebra, respectively.
To prove that there is no another symmetric Jacobi identity we
need few elementary algebraic concepts. By an {\it identity} of
the algebra $\calF$ we mean a polynomial $P$ in some free
algebra which is identically zero when the generators are
replaced by any elements of $\calF$.  We are interested on the
$3$-linear identities in which each term involves two pairs of
brackets, i.e. of the form: $a\prec \prec f,g \succ,h \succ +
b\prec \prec f,h \succ,g \succ + c\prec \prec g,h \succ,f
\succ=0$.  Now we would like prove the following result:
\begin{prop}
{\em There are exactly two types of $3$-linear identities for
     symmetric algebra in which each term involves two pairs of brackets:
     \begin{description}
     \item{{\bf a})} The {\it SE-identity} (symmetric version of the
          Engel identity):
          $\forall f \in \calF$: ~$\prec \prec f,f \succ,f \succ = 0$.
     \item{{\bf b})} The {\it SJ-identity}
          (symmetric version of the Jacobi identity):
          \beqnn
             \forall f,g : ~\prec \prec f,f \succ,g \succ =
                            \prec \prec f,g \succ,f \succ
             \Longleftrightarrow ~ \nonumber\\
             \forall f,g,h: ~
             2 \prec \prec f,g \succ,h \succ =
               \left[\prec \prec f,h \succ,g \succ +
                     \prec \prec g,h \succ,f \succ \right]  \,.
          \eeqnn
     \end{description}
}
\end{prop}
{\bf Proof.} ~ Suppose that $a\prec \prec f,g \succ,h \succ +
b\prec \prec f,h \succ,g \succ + c\prec \prec g,h \succ,f
\succ=0$. For $f=g=h$ we have $(a+b+c) \prec \prec f,f \succ,f
\succ = 0$, therefore the identity must be of the type {\bf a})
or $a+b+c=0$. If $a+b+c=0$, then for $f=h$ we have $(a+c)\prec
\prec f,g \succ,f \succ + b\prec \prec f,f \succ,g \succ=0$,
i.e. it must be of the type {\bf b}), since $a+c=-b$.
\begin{flushright}
{\bf Q.E.D.}
\end{flushright}
Similarly, one can prove that there are exactly two $3$-linear identities 
in which each term involves two pairs of brackets for antisymmetric 
algebra:
\begin{description}
     \item{{\bf a})} The {\it Engel identity}:
          \beqnn
               \forall f,g : ~\{\{f,g\},f\} = 0 ~
               \Longleftrightarrow \forall f,g,h :
               ~\{\{f,g\},h\} + \{\{h,g\},f\} = 0 \,.
          \eeqnn
     \item{{\bf b})} The {\it Jacobi identity}:
          \beqnn
               \forall f,g,h : ~
               \{\{f,g\},h\} + \{\{g,h\},f\} + \{\{h,f\},g\} = 0 \,.
               ~ \Box
          \eeqnn
\end{description}
Even the Jacobi identity and $SJ$-identity are formally similar
as we have seen above, but there exists also a fundamental
difference between them. Indeed, the former is equivalent to
$X_h \{f,g\} = \{ X_h f, g \} + \{ f, X_h g \}$, i.e.
Hamiltonian vector fields act as derivations for Poisson
bracket, but for the latter $ 2X_h^D \prec f,g \succ = \prec
X_h^D f, g  \succ + \prec  f, X_h^D g  \succ$.  In the other
words, the Jacobi identity manifests some basic features like
cyclicity, derivation and signature property while the
$SJ$-identity does not.

Furthermore, the most natural symmetric bracket $\prec A,B
\succ= \frac{AB+BA}{2}$ does not satisfy the $SJ$-identity in
the general. Hence, there is no natural representation for
$SP$-algebras.


\newpage


\end{document}